\documentclass{vietnam}

\def\be{\begin{equation}}
\def\ee{\end{equation}}
\def\bea{\begin{eqnarray}}
\def\eea{\end{eqnarray}}

\usepackage{amsmath,amssymb,mathtools,bm}
\usepackage{booktabs}
\usepackage{braket}
\usepackage{multirow}

\newcommand{\Photo}{}

\begin{document}
\vspace*{4cm}
\title{Dark Energy Driven by the Cohen--Kaplan--Nelson Bound}

\author{Patrick~Adolf${}^{a}$, Martin~Hirsch${}^{b}$, Sara~Krieg${}^{a}$, Heinrich~P\"as${}^{a}$, \underline{Mustafa~Tabet}${}^{a}$}

\address{${}^{a}$Fakult\"at f\"ur Physik, TU Dortmund, D-44221 Dortmund, Germany\\
${}^{b}$Instituto de F\`{i}sica Corpuscular, Universidad de Valencia-CSIC, E-46980 Valencia, Spain
}

\maketitle\abstracts{
In this work, we confront the bound on an ultraviolet cutoff (UV) of a quantum
field theory (QFT) proposed by Cohen, Kaplan, and Nelson (CKN) with the latest
results of the Dark Energy Spectroscopic Instrument (DESI). The former relates
the UV cutoff with an infrared (IR) cutoff of the theory by excluding all states
describing a black hole.
Identifying now the IR cutoff with the Hubble horizon yields
a time-varying contribution of the vacuum energy to the dark energy density of the
universe. At the same time the DESI results in combination with other cosmological data point
towards a preference of time-varying dark energy models over
$\Lambda$CDM.
}

\section{Introduction}
Quantum field theory (QFT) in its current form is incompatible with gravity.
This can be for example seen by introducing the concept of black hole entropy
in order to satisfy the second law of thermodynamics.
The latter is generalised in such a way that the sum of the black hole entropy and the
entropy exterior to the black hole never decreases~\cite{Bekenstein:1973ur,Bekenstein:1974ax,Bekenstein:1980jp,Bekenstein:1993dz,Hawking:1975vcx,Hawking:1976de}. The entropy of the black
hole necessarily scales with its horizon area~\cite{Bekenstein:1973ur,Bekenstein:1974ax,Bekenstein:1980jp,Bekenstein:1993dz,Hawking:1975vcx,Hawking:1976de}, while in a QFT the entropy scales
with the volume of the system. This implies that there is a length scale
characteristic to the quantum system where its entropy $S_\mathrm{QFT}$ in a box
of size $L$ exceeds the black hole entropy
$S_\mathrm{BH}$. As the latter sets an upper limit on the
entropy of any system the validity range of the QFT is constrained by the
inequality
\begin{align}\label{eq:bekenstein-bound}
    S_\mathrm{QFT} = \Lambda_\text{UV}^3 L^3 \leq \pi L^2 M_\mathrm{P}^2 = S_\mathrm{BH} \,,
\end{align}
where $\Lambda_\mathrm{UV}$ is the UV cutoff of the QFT, and $M_\mathrm{P}$ the
Planck mass.
The maximum length $L_\mathrm{IR}$ for which equation~\eqref{eq:bekenstein-bound}
is still satisfied, implies for a given UV (IR) cutoff $\Lambda_\mathrm{UV}$ of the
QFT a corresponding IR (UV) cutoff $\Lambda_\mathrm{IR} = 1/L_\mathrm{IR}$.
However, as Cohen, Kaplan, and Nelson (CKN) point out, there exist low-energy states
that can turn into a black hole while still satisfying equation~\eqref{eq:bekenstein-bound}~\cite{Cohen:1998zx}.
These states can be excluded by demanding the characteristic length $L_\mathrm{IR}$
to be larger than the Schwarzschildradius $R_\mathrm{S}$~\cite{Cohen:1998zx}
\begin{align}
    L_\mathrm{IR} \geq R_\mathrm{S} \sim \frac{M_\mathrm{QFT}}{M_\mathrm{P}^2}
        = \frac{\rho V}{M_\mathrm{P}^2}
        = \frac{\Lambda_\mathrm{UV}^4 L_\mathrm{IR}^3}{M_\mathrm{P}^2} \,,
\end{align}
with the energy density and volume $\rho$ and $V$ of the QFT, respectively. The
resulting UV cutoff now depends on the IR cutoff, the so-called CKN bound
\begin{align}\label{eq:ckn-bound}
    \Lambda_\mathrm{UV}^2 \leq \frac{M_\mathrm{P}}{L_\mathrm{IR}} \,.
\end{align}
The phenomenological consequences in particle physics have
been studied in several works~\cite{Cohen:2021zzr,Kephart:2022vfr,Adolf:2023wfw}. In this work, instead, we focus on the
cosmological consequence, namely that identifying the IR cutoff with the current
size of the Hubble horizon---as has already been done in the original
work~\cite{Cohen:1998zx}---implies a time-varying dark energy density. While this has been
done in the original work in order to show that the contributions of vacuum
integrals to the dark energy density are of the same order of magnitude as the
measurements, the time-dependence is particularly interesting in light of the
latest measurements of the Dark Energy Spectroscopic Instrument (DESI). These
measurements combined with other cosmological data prefer time-dependent dark
energy models with a statistical significance of up to $3.9\,\sigma$ compared
to $\Lambda$CDM~\cite{DESI:2024mwx}.

In section~\ref{sec:Evolving Dark Energy}, we first derive the cosmological
equations following from the CKN bound. The compatibility of the CKN bound with
the latest DESI results combined with model-independent Hubble measurements
and supernova data is presented in section~\ref{sec:Analysis and Results}.
Finally, we summarise in section~\ref{sec:Summary and Outlook}.

\section{Evolving Dark Energy}
\label{sec:Evolving Dark Energy}
The CKN bound in equation~\eqref{eq:ckn-bound} with the Hubble horizon as an IR cutoff
implies
\begin{align}
    \Lambda^4_\mathrm{UV} \lesssim H^2(z) M_\mathrm{P}^2 \,.
\end{align}
The vacuum energy density $\rho_\mathrm{VED}^\text{1-loop}$ is given by summing the harmonic
oscillator modes between the resulting IR and UV cutoff
\begin{align}\label{eq:1-loop-ved}
    \rho_\mathrm{VED}^\text{1-loop}
    \simeq \int_{\Lambda_\mathrm{IR}}^{\Lambda_\mathrm{UV}}
        \frac{4\pi k^2\, \mathrm{d}k}{\left(2\pi\right)^3} \sqrt{k^2 + m^2}
    \simeq \frac{\Lambda^4_\mathrm{UV}}{16\pi^2}
    \simeq \nu \frac{H^2(z) M_\mathrm{P}}{16\pi^2} \,.
\end{align}
Note that here, we neglect the contribution coming from the IR cutoff,
$\Lambda_\mathrm{IR}^4 \propto H^4(z)$, as we are analysing data where $H^4(z) \ll M_\mathrm{P}^2H^2(z)$.
Note that in equation~\eqref{eq:1-loop-ved}, we further introduce a prefactor $\nu$
as the actual value of the loop integral depends for example on the particles contributing
between the IR and UV cutoff, and thus, on the underlying particle physics model.
This prefactor also has the advantage of parameterising other models in the literature where
the vacuum energy density also scales proportional to the Hubble parameter~\cite{Abdalla:2022yfr}.
The correction resulting from equation~\eqref{eq:1-loop-ved} is now incorporated semi-classically
into the energy-momentum tensor $T^{\mu\nu}$
\begin{align}
    T^{\mu\nu}_\mathrm{tot} = T^{\mu\nu}_\mathrm{classical} + \rho_\mathrm{VED}^\text{1-loop} g^{\mu\nu} \,,
\end{align}
with
\begin{align}\label{eq:emt-conservation}
    \nabla_\mu T^{\mu\nu}_\mathrm{tot} = 0 \,.
\end{align}
Note that, we assume the conservation of the total energy momentum tensor instead
of the conservation of each contribution to it. Otherwise, it immediately follows
from equation~\eqref{eq:emt-conservation} that the time-dependence of the vacuum energy
density needs to vanish, i.e. $\dot\rho_\mathrm{VED}^\text{1-loop} = 0$.
Moreover, a scaling like the one in equation~\eqref{eq:1-loop-ved} would not be
able to explain the accelerated expansion of the universe~\cite{Hsu:2004ri,Li:2004rb}. Focusing on
the matter dominated era and solving equation~\eqref{eq:emt-conservation}
together with the usual Friedmann equation
\begin{align}
    H^2(t) = \frac{8\pi G}{3}\left( \rho_\mathrm{M}(t) + \rho_\Lambda(t) \right) \,,
\end{align}
yields the matter and dark energy densities
\begin{align}
    \Omega_\mathrm{M}(z) &= \Omega_M^0 (1 + z)^{3 - \frac{\nu}{2\pi}} \,,\\
    \Omega_\Lambda(z)    &= \Omega_\Lambda^0 + \Omega_M^0 \frac{\nu}{6\pi - \nu}\left[
        (1 + z)^{3 - \frac{\nu}{2\pi} - 1}
    \right] \,,
\end{align}
where $\Omega_M^0$ and $\Omega_\Lambda^0$ denote the matter and dark energy density today
normalized to $\rho_{\mathrm{crit}, 0} = 3H_0^2/(8\pi G)$, respectively.

\section{Analysis and Results}
\label{sec:Analysis and Results}
Here, we perform an analysis to determine the agreement between the CKN bound
and current experimental constraints. We focus on the recent DESI
baryonic acoustic oscillations (BAO) measurements~\cite{DESI:2024mwx} combined with late universe
constraints, in particular, we use the supernova distance datasets of DES-SN5YR (DESY5)~\cite{DES:2024tys}
and Pantheon+~\cite{Brout:2022vxf}, and model-independent Hubble parameter measurements~\cite{Favale:2024lgp}.
Due to the overlap between the supernova distance datasets, we focus on one dataset at a time.
The agreement between those experimental measurements with the theory prediction is quantified
using $\chi^2$-statistics
\begin{align}
    \chi^2 = \left( \vec{O}_\mathrm{th}(\xi) - \vec{O}_\mathrm{exp} \right)^T
        C^{-1} \left( \vec{O}_\mathrm{th}(\xi) - \vec{O}_\mathrm{exp} \right) \,,
\end{align}
where the vector $\vec{O}_\mathrm{th}$ denotes the theory predictions as a
function of the model parameters $\xi_i$, while the corresponding measurements
with the covariance matrix $C$ are denoted by the vector $\vec{O}_\mathrm{exp}$.

\begin{figure}
    \centering
    \includegraphics[width=0.3\textwidth]{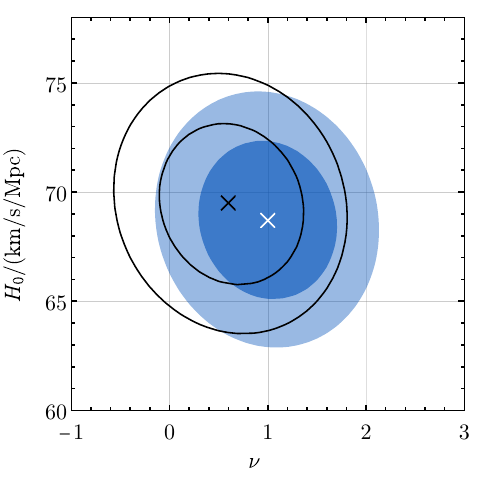}
    \includegraphics[width=0.3\textwidth]{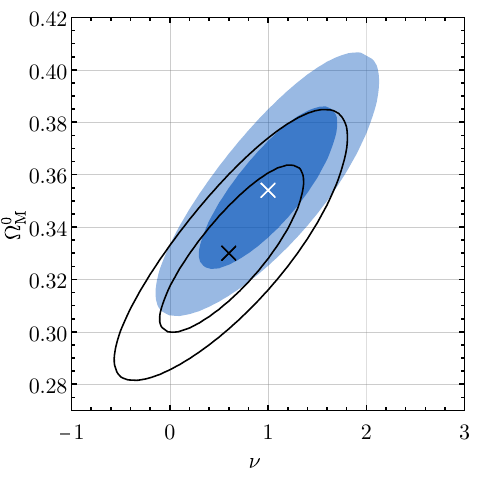}
    \includegraphics[width=0.3\textwidth]{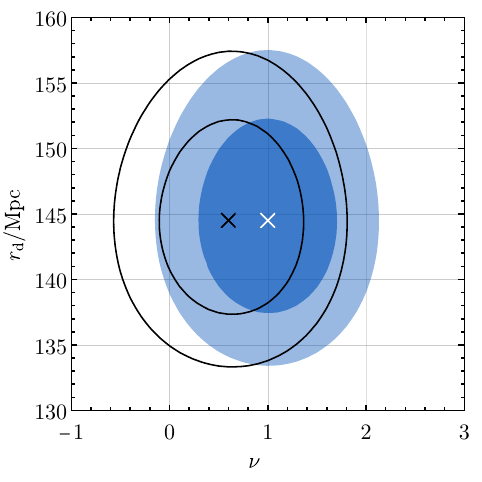}
    \caption{Allowed region in the $\nu$--$H_0$, $\Omega_M$, $r_d$ plane
        at $68\,\%$ and $95\,\%$ CL for the combination with the DESY5 (blue)
        and Pantheon+ (black) dataset~\protect\cite{Adolf:2024twn}.}
    \label{fig:correlations}
\end{figure}

\begin{figure}
    \centering
    \includegraphics[width=0.4\textwidth]{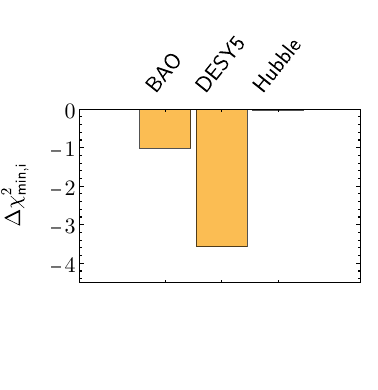}
    \quad
    \includegraphics[width=0.4\textwidth]{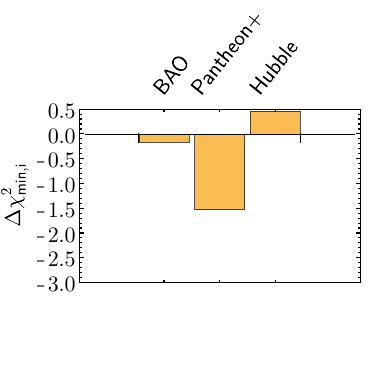}
    \vspace{-50pt}
    \caption{The contribution of each observable to the difference in the minima of the $\chi^2$
        distributions between $\nu$CKN and the $\Lambda$CDM model.}
    \label{fig:pulls}
\end{figure}

Combining the DESI BAO and Hubble parameter measurements with either the DESY5 or Pantheon+
dataset, respectively, yields a minimum of $\chi^2$ of
\begin{align}
    \chi^{2,\mathrm{\nu CKN}}_\mathrm{min, \,DESY5}/\mathrm{dof} &= 1677/1870 \approx 0.90 \,,
    &&\chi^{2,\mathrm{\nu CKN}}_\mathrm{min, \,Pantheon+}/\mathrm{dof} = 1440/1632 \approx 0.88 \,.
\end{align}
The allowed parameter space for the prefactor $\nu$ and the
remaining model parameters is shown in figure~\ref{fig:correlations}.
It is also interesting to see how this compares with the well-established $\Lambda$CDM model.
The difference in the minima of the $\chi^2$ distributions of both models reads
\begin{align}
    \Delta \chi^2_\mathrm{min, \,DESY5} &= -4.6 \,,
    && \Delta \chi^2_\mathrm{min, \,Pantheon+} = -1.1 \,,
\end{align}
for the DESY5 and Pantheon+ datasets, respectively, which translates to a preference
of $2.1\,\sigma$ and $1.1\,\sigma$ of the $\nu$CKN case compared to $\Lambda$CDM.
The contribution of each observable to the difference in the minima of the $\chi^2$ distributions
is shown in figure~\ref{fig:pulls}.
It is also interesting to compare the $\nu$CKN case with other models of time-varying dark energy.
This has been done in reference~\cite{Adolf:2024twn}, and we refrain from presenting the results again.

\subsection*{Future Projection}
DESI will run for additional three years; and other experiments like Large Synoptic
Survey Telescope or Euclid are either planned or already have started collecting
data last year. This will result in significantly better statistics, and therefore,
smaller uncertainties which allows for more significant discrimination between
various cosmological models.

In anticipation of these new data, we perform for the first time a future projection by
assuming that the central value of the current measurements will stay the same but
with reduced uncertainties. In the case of DESI, we reduce the uncertainties
on the year-1 DESI results by a factor of $\sqrt{5}$ to account
for the total data taking period of five years. To account for the increased
statistics in the supernova distance datasets, we reduce the uncertainty by a factor
of 4 which corresponds to the lowest expected improvement on the distance luminosity
measurements in the case of Euclid. This projection is referred to as ``Euclid-Unc''
while we refer to the DESI projection as ``DESI-5Y''.

In table~\ref{tab:fp}, we show the discriminatory power of these future projection, as an example
for the comparison between the $\nu$CKN and $\Lambda$CDM models. The comparison with other
different models can again be found in reference~\cite{Adolf:2024twn}.
\begin{table}
    \begin{center}
    \caption{Shown is the future projection of the preferences of the DESI experiment after five years of measurements
        (DESI-5Y) and the Euclid-Unc experiment between the $\nu$CKN and $\Lambda$CDM~\protect\cite{Adolf:2024twn}.}
    \begin{tabular}{c c c c c c c}
    \toprule
    \multirow{2}{*}{\bf Models} & \multicolumn{2}{c}{DESI-5Y} & \multicolumn{2}{c}{Euclid-Unc}& \multicolumn{2}{c}{DESI-5Y + Euclid-Unc}\\
    \cmidrule(l){2-3}
    \cmidrule(l){4-5}
    \cmidrule(l){6-7}
    & $\Sigma_\text{DESY5}$ & $\Sigma_\text{Pantheon+}$ & $\Sigma_\text{DESY5}$  & $\Sigma_\text{Pantheon+}$ & $\Sigma_\text{DESY5}$  & $\Sigma_\text{Pantheon+}$\\
    \midrule
    \textbf{$\bm{\nu}$CKN with}\\
    $\Lambda$CDM             & $-1.9\sigma$     & $-1.0\sigma$    & $-4.3\sigma$  &   $-3.1\sigma$  & $-6.6\sigma$   & $-4.4\sigma$ \\
    \bottomrule
    \end{tabular}
    \label{tab:fp}
    \end{center}
\end{table}

\section{Summary and Outlook}
\label{sec:Summary and Outlook}
The latest DESI BAO data combined with late universe constraints, namely,
the two supernova datasets, DES-SN5YR (DESY5) and Pantheon+, and model-independent
Hubble parameter measurements are well compatible with the CKN bound.
Interestingly, the ($\nu$)CKN model provides a better fit to the data than $\Lambda$CDM for both
supernova datasets considered in this work. However, while the preference for the DESY5
datasets is around $2.1\,\sigma$, the preference using the Pantheon+ dataset is around
$1.1\,\sigma$, and thus, non-significant. For further comparisons between different models
see reference~\cite{Adolf:2024twn}.
A rough projection of the improved statistics expected in the near future shows that one might be
able to distinguish between a QFT respecting the CKN bound where the vacuum contributions
contribute to the dark energy density of the universe and $\Lambda$CDM with a statistical
significance with up to $6.6\,\sigma$.

\section*{Acknowledgments}
This contribution to the Dark Matter and Dark Energy parallel session of PASCOS 2024 is
based on reference~\cite{Adolf:2024twn}.

\end{document}